\documentclass[11pt]{iopart}

\usepackage{graphicx}% Include figure files
\usepackage{dcolumn}% Align table columns on decimal point
\usepackage{bm}% bold math
\usepackage{amsfonts}
\usepackage{amssymb}

\def\be{\begin{equation}}
\def\ee{\end{equation}}
\def\ben{\begin{displaymath}}
\def\een{\end{displaymath}}
\def\ba{\begin{eqnarray}}
\def\ea{\end{eqnarray}}
\def\nn{\nonumber}
\def\os{Ozsv\'{a}th and Sch\"ucking }
\def\osa{Ozsv\'{a}th-Sch\"ucking }

\def\oz{Ozsv\'{a}th}

\def\arsinh{\mathop{\rm arsinh}\nolimits}

 \renewcommand{\bf}{\bfseries}
 \renewcommand{\it}{\itshape}
\begin{document}
                         
\title[G\"odel-like lightcones]{The lightcone of 
G\"odel-like spacetimes}

 \author{G.~Dautcourt\footnote{  
Email:~daut@aei.mpg.de,
MPI f\"ur Gravitationsphysik, 14476 Golm, Germany}}

%\address{Max Planck Institut f\"{u}r Gravitationsphysik,
%Albert-Einstein-Institut,\\ 
%Am M\"{u}hlenberg 1, D-14476 Golm, Germany\\
\date{\today}

\begin{abstract} 
A study of the lightcone of the G\"odel universe is extended 
to the so-called G\"odel-like spacetimes. 
This family of highly symmetric 4-D Lorentzian spaces is defined 
by metrics of the form  
$ds^2=-(dt+H(x)dy)^2+D^2(x)dy^2+dx^2+dz^2$,
together with the requirement of spacetime homogeneity,  
and includes the G\"odel metric.  The quasi-periodic refocussing
of cone generators with startling lens properties, discovered by
Ozsv\'{a}th and Sch\"ucking for the lightcone of a plane gravitational 
wave and also found in the G\"odel universe,
is a feature of the whole G\"odel family.  
We discuss geometrical properties of caustics and show that 
(a) the focal surfaces are two-dimensional null surfaces generated by 
non-geodesic null curves and (b) intrinsic differential invariants of the 
cone attain finite values at caustic subsets.                      
\end{abstract}
\date{\today}

\section{Introduction}

The study of null congruences, null hypersurfaces
and in particular lightcones in general relativity
is  complicated by the existence of caustics, i.e.  
points of intersection of the generating null geodesics.    
Since light rays become focussed in the presence of matter as well as
shear, caustics occur inevitably in realistic situations as frequently 
encountered in astrophysics. 
The corresponding strong lens effect  
is an important astrophysical tool                                      
\cite{schn92}. 
In numerical relativity caustics are less welcome, they act as     
barrier for current characteristic codes \cite{friedr83,
cork83,winic98}.
The differential geometry of caustics in a spacetime setting is still        
not well developed,
contrary to their mathematical classification using methods of 
singularity theory \cite{arnold85}. This may be due to the fairly 
complicated structure of these objects, involving crossings and singularities.
Important steps have been taken, among others,                             
 by Friedrich and Stewart \cite{friedr83}  and by the Newman school 
\cite{frittn1}-\cite{fritt03}, \cite{kling}.

A way towards a better understanding is the study 
of curved spacetimes with analytically known focal surfaces.
\os presented already in 1962 an exact and detailed analytical  
picture of the lightcone of a plane gravitational wave \cite{os62}.
They found a cyclic structure of the focal set, produced by a semi-periodic 
re-focussing of light rays. An often reproduced illustration of a similar light 
cone drawn by Penrose \cite{pen65} served as starting point for investigations 
in global Lorentzian geometry \cite{ehr}, \cite{beem}. 
A very similar focal structure is present on the lightcone of a quite 
different spacetime, the rotating G\"odel universe 
\cite{kundt56,abdel72b,da06}. 
In view of the T-duality  
of higher-dimensional supersymmetric versions of the G\"odel metric
and pp waves \cite{boyda03,harmark}  it is perhaps not surprising  
that the same type of caustic is present.
         
For further analytical studies of caustics it makes sense  
to discuss spaces of high symmetry first, 
since here the geodesic equations can be integrated completely.          
Furthermore, if spacetime homogeneity applies,
all lightcones have the same intrinsic geometry, independent of the vertex
location. 
To this class of spacetimes belong metrics of the \oz~ class III \cite{os70}, 
which include the G\"odel metric. They  have been studied by Rosquist et al
\cite{laur81}.     
Other examples are 
the spacetime-homogeneous G\"odel-like or G\"odel type  
metrics  \cite{som68},\cite{RG},\cite{CB},\cite{RT1}-\cite{RT3},\cite{TRA},
\cite{CRTS}, also generalizations of the G\"odel metric and
admitting at least a $G_5$ Killing symmetry.
Their lightcone, and in particular the focal subset, 
is the subject of the present paper. 

Section 2 shortly reviews 
the two-parameter family of G\"odel-like metrics. 
Basic geometrical 
properties of these metrics depend on a dimensionless parameter $k$, 
measuring the influence of rotation on the spacetime geometry.
$k^2$ may range from 
$-\infty$ to $\infty$, but in this article we confine the discussion to 
a range of positive $k^2$. 
The  $k$-sequence coincides with the family 
of $(2+1)-$dimensional geometries investigated by Rooman and Spindel 
\cite{rooman98}, if one flat space dimension is added to Rooman-Spindel.  
Their parameter $\mu$ is our $k$. 

The lightcone geometry of the G\"odel family with 
$1<k^2<\infty$ ($k^2=2$ corresponds to the G\"odel cosmos \cite{godel49},
\cite{godel52}, \cite{hawell73}, \cite{barrow}, \cite{os01}, \cite{os03}) 
 is studied in section 3, based on a paper by 
Calv\~{a}o, Soares and Tiomno \cite{CST}. Further sections consider 
briefly some limiting cases. 
In section 4 we treat the special case $k^2=1$, known as Rebou\c{c}as-Tiomno 
metric \cite{RT1,RT2}.
Its subspace $z=const$ is the three-dimensional anti-deSitter space $AdS_3$.
The causal family $0<k^2<1$ without
closed timelike curves is omitted here, only 
the static degeneration $k^2 \rightarrow 0$  with 
vanishing rotation is shortly considered in section 5.
The concluding section notes that a cyclic behaviour of caustics             
on many lightcones (and on null hypersurfaces in general) may be expected 
as a consequence of the Sachs equations \cite{sachs61} for 
divergence and shear of the generator congruence. 

\section {G\"odel-like metrics}

Raychaudhuri and Guha Thakurta \cite{RG} have introduced as "homogeneous 
spacetimes of the G\"odel type" the metrics
\be
ds^2 = -(dt + H(x)dy)^2 +D^2(x)dy^2+dx^2+dz^2, \label{gt0}
\ee
together with the additional condition of spacetime homogeneity. Spacetime 
homogeneity requires at least one further Killing field, 
apart from the three translational Killing vectors along the axes, 
which evidently exist.
This leads to the {\it necessary} conditions 
\be
H'/D = const =2\Omega,~~D''/D = const =l^2, \label{sth}
\ee
with two parameter $\Omega, l$.  We replace $\Omega$ by the    
parameter $k= 2\Omega/l$ and consider the sequence of metrics labeled by 
$k$. Only real $k$ are taken here and $l$ is assumed non-negative.
Rebou\c{c}as and Tiomno found that the conditions (\ref{sth}) are 
also {\it sufficient} for spacetime homogeneity \cite{RT1}. Moreover, 
it was shown
in \cite{RT1} that for the metrics satisfying (\ref{gt0}) 
and (\ref{sth}) a further Killing vector exists, leading to a $G_5$ group 
of motions. A re-examination of the symmetries by Teixeira, Rebou\c{c}as and 
{\AA}man \cite{TRA}, who dropped the more or less implicit assumption of 
time-independent Killing fields made so far, has shown that in the special 
case $k^2=1$ the group is $G_7$, the maximal symmetry group 
within the G\"odel-like class of spacetimes.

We write the metric (\ref{gt0}) in cylindrical coordinates 
$(t,r,\phi,z)$ as 
\be
ds^2 = -(dt +\frac{2k}{l}\sinh^2{(lr/2)}d\phi)^2
+ \frac{\sinh^2{(lr)}}{l^2} d\phi^2 +dr^2+dz^2. \label{gt1}
\ee
The numbers $(k^2,l)$ in the two-parameter family (\ref{gt1})
specify a metric uniquely: Members with different pairs
$(k^2,l)$ represent different spacetimes. 
In the limit $k\rightarrow\infty$ and $l\rightarrow 0$ such that      
$kl=2\Omega$ remains finite, the function 
$\sinh{(lr)}/l$ can be replaced by $r$, and (\ref{gt1}) becomes the 
Som-Raychaudhuri metric \cite{som68}. For $k^2 \rightarrow 2$ one 
recovers the  G\"odel metric                                      
and for $k^2=1$  the already mentioned $G_7$ metric is obtained, 
studied in detail by Rebou\c{c}as and Tiomno \cite{RT1}.

For convenience we note some properties of the metric (\ref{gt1}).
The non-vanishing components of the Ricci tensor are (we follow the
conventions of \cite{exact}) 
\be \fl \qquad
R_0^{~0} = -k^2l^2/2,~ 
R_1^{~1} =  R_2^{~2} =  l^2(k^2-2)/2,~                            
R_2^{~0} =  kl (k^2-1)(1-\cosh{(lr)}).    \label{ricci}       
\ee
The eigenvalues $\lambda$ determined from 
$det |R_\mu^{~\nu}- \lambda \delta_\mu^{~\nu}| =0$    
 follow as                   
\be
\lambda_1 =0,~\lambda_2=-k^2l^2/2,~\lambda_{3,4}=l^2(k^2-2)/2.
\ee 
The Weyltensor, also given in coordinate form, has the non-vanishing components
\ba    \fl \qquad
C_{0101} &=& -C_{1313} =(k^2-1)l^2/6,~~C_{0112} =-kl(k^2-1)\cosh{(lr)}/6, \nn \\
  \fl \qquad
C_{0202} &=& (k^2-1)\sinh{(lr)}/6,~~C_{0303} =  -(k^2-1)l^2/3, \nn  \\ 
  \fl \qquad
C_{0323} &=& -kl(k^2-1)(\cosh{(lr)}-1)/3,   \label{weyl} \\          
  \fl \qquad
C_{1212} &=& (k^2-1)(\cosh{(lr)}-1)(k^2(\cosh{(lr)}-1)+2\cosh{(lr)}+2)/6, \nn \\
  \fl \qquad
C_{2323} &=&(k^2-1)(\cosh{(lr)}-1)(-2k^2(\cosh{(lr)}-1)-1-\cosh{(lr)})/6. \nn   
\ea 

There exist several interpretations of the matter tensor as calculated
from the Einstein field equations (including a cosmological constant).
The $k \rightarrow   \infty$ limit, the Som-Raychaudhuri metric, 
describes the gravitational field of a homogeneous distribution of charged 
rotating dust.
For the G\"odel family $1 < k^2 <\infty$, the
combination of a perfect fluid, a scalar field and a homogeneous 
source-free electromagnetic field may serve as matter                  
\cite{RT1}. A perfect
fluid description alone applies only to the G\"odel metric $k^2=2$, 
as shown by Bampi and Zordan \cite{bampi78}. For more recent discussions 
of G\"odel-like 
spacetimes in gravity theories derived from Lagrangians which are arbitrary
functions of curvature invariants, see \cite{clift},\cite{RS}. 

The interpretation of the G\"odel family as solutions of the Einstein or 
other field equations may be considered as dubious 
in the sense that unusual or unphysical forms of matter are involved.
Therefore these metrics (except G\"odel) are not treated in the standard book
on exact solutions \cite{exact}. 
But this aspect is not important for the geometrical discussion in this 
paper. The metrics are mainly interesting for their {\it high degree of 
symmetry}. 
All admit at least a $G_5$ group of motions.

\section{The G\"odel family $ 1<k^2<\infty$}
\subsection{Null geodesics}

For generic G\"odel-like metrics, 
use is made of the results by Calv\~{a}o, Soares and 
Tiomno (CST) \cite{CST}, also largely keeping their notation for comparison. 
The authors follow a previous paper by Novello, Soares 
and Tiomno dealing with the G\"odel metric \cite{NST}. They 
give a complete discussion of timelike geodesics and treat also
null geodesics. We consider only the lightlike case.        
The high symmetry of the metric allows
to write down a sufficient number of first 
integrals for the geodesic equations 
$Dx^\mu/Ds^2=0$, using the fact that for a Killing field $k_\mu$, 
$\frac{dx^\mu}{ds} k_\mu$ is constant 
along a geodesic. With the Killing translations 
 $\partial_t,~\partial_\phi,~\partial_z$ and the corresponding integration 
constants $p_t,~p_\phi,~p_z$ or equivalently $p_t,~\beta= p_z/p_t,~
\gamma=p_\phi/p_t$, three first integrals may be written  
\ba
\dot{t}/p_t &=&  1+ kl/2 -k^2\sinh^2{(lr/2)}/\cosh^2{(lr/2)},\label{tdot}  \\
\dot{\phi}/p_t &=& kl/(2\cosh^2{(lr/2)})  
-l^2\gamma /(4\sinh^2{(lr/2)}\cosh^2{(lr/2)}), \label{Phdot}\\
\dot{z}/p_t &=& -\beta. \label{zdot}
\ea
The dot denotes the derivative with respect to an affine parameter $s$.
A further relation follows from 
$\frac{dx^\mu}{ds}\frac{dx^\nu}{ds}g_{\mu\nu}=0$:
\be
\dot{r}^2/p_t^2 = 1-\beta^2 
-\Bigl(k\frac{\sinh{(lr/2)}}{\cosh{(lr/2)}} 
-\frac{l\gamma}{2\sinh{(lr/2)}\cosh{(lr/2)}}\Bigr)^2.\label{rdot}
\ee 
As shown by CST, it is convenient to introduce instead of
 $r$ another radial variable $\xi$, which increases  monotonically with $r$:
\be
\xi = \sinh^2{(lr/2)}. \label{xi}
\ee 
Equation (\ref{rdot}) then becomes
\be
\dot{\xi}^2/p_t^2  = l^2\eta\xi^2 +l^2(1-\beta^2
+ k l \gamma)\xi -l^2\gamma^2/4 \label{xid}
\ee
with
\be
\eta =k^2+\beta^2-1.
\ee
The equations (\ref{tdot})-(\ref{xid}) refer to the class of {\it all}  
null geodesics. We are interested in the subset forming a single cone, e.g., 
passing through the origin of the coordinate system, $t=0,~r=0,~z=0$.         
This subset is obtained by setting $\gamma= 0$: Expanding the 
rhs of (\ref{xid}) around $r=0$ or $\xi \approx l^2r^2/4 =0, $ one obtains
$(d\xi/ds)^2 \approx -l^4 p_t^2\gamma^2/4  <0 $, hence no geodesics 
with $\gamma \neq0$ can pass the origin. On the other hand, 
every geodesic with $\gamma=0$ passes the origin.         

With $\gamma=0$ the first integrals simplify considerably. 
The geodesic equations can be integrated completely, leading to the following 
 parameter representation of the light
cone with vertex at $t=0,~r=0,~z=0$:

\ba
t &=&  \frac{2k}{l}\arctan{\Bigl(\frac{k}{\sqrt{\eta}}
\tan{w}\Bigr)}  
- \frac{2w(k^2-1)}{l\sqrt{\eta}},  \label{teq}\\ 
r &=& \frac{2\epsilon}{l} \arsinh{\Bigl(\frac{\sin{w} \sqrt{1-\beta^2}}
{\sqrt{\eta}}\Bigr)},   \label{req}\\
\phi &=& \phi_0 
+  \arccos{(\sqrt{1-\beta^2-\eta \xi}/(\sqrt{1-\beta^2}\sqrt{1+\xi}) )},    
\label{phieq}\\  
z &=& -\frac{2w\beta}{ l\sqrt{\eta}}, \label{zeq}
\ea
with $\epsilon =1 (-1)$ for the future (past) cone.
We have introduced a new affine parameter $w$ 
instead of $s$ by

\be 
        w =lp_t\sqrt{\eta}(s-s_0)/2   \label{affpar} 
\ee
(note $\eta>0$, since $k^2>1$ is assumed, the case $k^2=1$ is treated 
separately). $w >0 (< 0)$ corresponds to the future (past) cone. 
 Equation (\ref{phieq}) differs from the corresponding equation (47)
- restricted to  $\gamma=0$ - in \cite{CST}. Both equations are correct,
but refer to different initial values $\phi_0$.
We have replaced the CST equation in order to have $\phi=\phi_0$ at the 
origin $r=0$. 

The cone generators depend on the two parameter $\beta$ and $\phi_0$,
which represent a possible pair of transversal coordinates for the 
light rays.
It appears more useful to introduce (primarily for the past cone, but 
easily extended to the full cone) the two angular 
coordinates $\theta, \varphi$ on the sky of a suitable observer at the vertex. 
The observer is assumed comoving with the cosmic fluid  with the timelike 
velocity vector $u^\mu= \delta^\mu_0 $ (in the case of the G\"odel metric 
with $k^2=2$) or defined by the normed timelike eigenvector of the Ricci tensor in general.           
It then is not difficult to see (e.g. by using the method described in the 
second Appendix in \cite{da06}) that $\beta$ and $\phi_0$ 
are related to the coordinates $\theta, \varphi$ (polar angle and longitude) 
on the observer sky by

\be
   \beta = \cos{\theta},~~ \phi_0 = \varphi. \label{skyc}
\ee
We note some well-known or easily accessible results.
From (\ref{req}) follows that null geodesics from the origin  
re-converge after reaching (for rays labeled $\theta$)  a maximal 
radial extension $r_{\theta}$, so we always have 
\be
r \leq r_{\theta}=
\frac{2}{l}\arsinh{\Bigl(\frac{\sin{\theta}}{\sqrt{k^2-\sin^2{\theta}}}\Bigr)}. 
\ee 
$r_{\theta}$ is zero for
rays along the polar    axis and in the opposite (antipode) direction
($\theta=0,\pi$) and reaches its largest value $r_m$ 
for equatorial rays ($\theta=\pi/2$). The hypersurfaces $r=const$ are always 
timelike, in particular  $r=r_m$ is the so-called "light cylinder" or 
"optical horizon": 
Evidently, the spacetime region $r>r_m$ 
cannot be reached by null geodesics from the origin. 

At the horizon $r=r_{m}$ the coefficient of $d\phi^2$ in (\ref{gt1}) is zero, 
thus the $\phi$-coordinate lines become  closed lightlike 
(for $r>r_{max}$, timelike) lines. They are not geodesics, however. 
A theory of {\it non-geodesic null curves} in a  Minkowski spacetime
was developed by Bonnor \cite{bonnor69}. His approach translates
immediately to curved spacetimes.
A calculation shows that the closed null curves on the optical 
horizon are {\it null helices} with constant Bonnor curvatures 
$k_1=1, k_2=l(1+k^2)/(4k), k_3=0$.
                     
\subsection{Focal subsets and inner metric}
 
The equations (\ref{teq})-(\ref{zeq}) supplemented by (\ref{skyc}) map 
the intrinsic coordinates $(w,\theta,\varphi)$ of the lightcone 
to the spacetime coordinates $(t,r,\phi,z)$.  
The critical points of this map are those where the Jacobian matrix 
has not the maximal rank 3. This happens if close cone generators intersect. 
For the critial or focal points all four subdeterminants of the Jacobian 
must vanish simultaneously: 
\be
 \frac{\partial (r,\phi,z)}{\partial (w,\theta,\varphi)}=0,~
 \frac{\partial (\phi,z,t)}{\partial (w,\theta,\varphi)}=0,~
 \frac{\partial (z,t,r)}{\partial (w,\theta,\varphi)}=0,~
 \frac{\partial (t,r,\phi)}{\partial (w,\theta,\varphi)}=0.\label{foc} 
\ee 
A straightforward calculation of (\ref{foc}) 
leads to the condition $f(w,\theta)=0$ for the focal set,  where
\be
f(w,\theta) \equiv  k^2\sin{w}\cos^2{\theta}                               
+ (k^2-1)w \cos{w} \sin^2{\theta}.                       
\ee

Another way to find singularities is to look for higher 
degeneration of the induced lightcone metric.
Numbering the inner coordinates as $y^1=w,~y^2=\theta,~ y^3=\varphi$, 
the intrinsic three-dimensional cone metric is determined by ($i,k=1...3$)
\be
\gamma_{ik} = \frac{\partial x^\mu}{\partial y^i}
\frac{\partial x^\nu}{\partial y^k}g_{\mu\nu}.
\ee
$\gamma_{ik}$ is already degenerate of rank 2. 
A direct calculation shows that the only
nonvanishing independent components are
\ba
\fl \qquad \gamma_{22} &=&\frac{4(f^2 -2 f k^2 \cos^2{\theta}\sin^3{w} 
+k^2q\cos^2{\theta}\sin^4{w})} {l^2\eta^3\cos^2{w}}, \label{m22}\\
\fl \qquad \gamma_{23} &=& 
\frac{4k\sin{\theta}\cos{\theta}\sin^2{w} (f-q\sin{w})}
{l^2\eta^{5/2}\cos{w}},  \label{m23} \\
\fl \qquad \gamma_{33} &=& \frac{4q\sin^2{\theta}\sin^2{w}}{l^2\eta^2}.
 \label{m33}
\ea
To obtain these compact expressions                         
we have introduced - besides the focal function $f(w,\theta)$ - 
  a non-negative function $q(w,\theta)$:
\be
q(w,\theta)  = (k^2+\cos^2{\theta} -1) \cos^2{w} +k^2\cos^2{\theta}\sin^2{w}. 
\ee
For later use we note that $q$ is zero on some closed $\varphi$-coordinate
lines, defined by $\theta=\pi/2$,     
$w=(2n-1)\pi/2, n=0,\pm 1,\pm 2,\pm 3...$ and arbitrary $\varphi$ in the 
range ($0,2\pi$).
          
The determinant of the two-dimensional metric (\ref{m22})-(\ref{m33}) can be 
written as square of a function $h(w,\theta)$:
\ba
 \gamma_{22}\gamma_{33}-\gamma_{23}^2=h^2,    \\
 h(w,\theta) = 4f(w,\theta)\sin{\theta} \sin{w}/(l^2\eta^2).
\label{heq} 
\ea
Higher order degeneration of the metric requires $h=0$ and is therefore 
given by  
 
(i) the set of focal points  $f(w,\theta)=0$, where neighbouring light rays
   intersect,

(ii) the set of points with 
$ w=n\pi$, $n$ integer, called "keel" singularities in  
\cite{da06}, where all rays with equal $\theta $ and different $\varphi$  meet 
in a point on the $n$th keel, a spacelike line of finite length, and

(iii) the pole rays $\theta = 0, \pi$,  
resulting from the $\sin{\theta}$-factor, i.e. from the singularity of
the polar coordinate system.

\subsection{Newman-Penrose coefficients on the cone}

Additionally to the intrinsic metric, 
the geometry of a null hypersurface may be described by some of the 
Newman-Penrose spin coefficients, mainly by 
divergence and shear and their change along a ray.

To illustrate this we first shortly consider                                 
a fairly known example, a generic null hypersurface in a Minkowski 
spacetime. Here the real divergence and complex shear evolve along a given 
ray according to the Penrose equations \cite{pen61}
\be
\rho =      (\rho_0+w[\sigma_0\bar{\sigma_0}-\rho_0^2]     )/f,~~
\sigma= \sigma_0/f \label{mink}
\ee
with the focal function 
\be
f = 1-2w\rho_0+w^2(\rho_0^2-\sigma_0\bar{\sigma_0}).
\ee
$\rho_0$ and $\sigma_0$ depend on the two transversal parameter
fixing a ray. From (\ref{mink}) we have                          
\be
\rho^2-|\sigma|^2= (\rho_0^2-\sigma_0\bar{\sigma_0})/f. \label{mpar}
\ee 
This equation shows that a parabolic point (a point with  
$\rho^2=|\sigma|^2 $) on a Minkowskian ray implies that 
the whole ray consists of parabolic points - provided 
the denominator $f$ in (\ref{mpar}) does not vanish.                  
The denominator vanishes and thus both $\rho$ and $|\sigma|$ diverge,
if the affine parameter $w$ takes one of the two values             
\be
  w_f= 1/(\rho_0 \pm |\sigma_0|).
\ee
At each $w_f$ a focal surface is passed, and the sign of 
 $\rho^2-|\sigma|^2$ changes.            
The quotient $j=\rho/|\sigma|$
remains finite, more exactly, 
$j \rightarrow \pm 1$ at a focal point. One also notes that              
a focal point can be considered as degenerate parabolic point.  

A similar behaviour of the first-order invariant $j$ at caustics 
holds for the \osa plane wave lightcone \cite{abdel72a}
and was found in \cite{da06}  for the lightcone of the G\"odel metric.
The difference is only that in both cases one meets an {\it unlimited} number 
of focal points if one moves along a ray. It is easy to see that this holds   
for G\"odel-like metrics in general:
If one starts from a                                 
cone metric $\gamma_{AB}$ ($A,B$ always run $2,3$) with 
$h=\sqrt{det|\gamma_{AB}|}$ and  $w$ as running (not necessarily affine) 
 parameter on the generating 
rays, divergence and shear amount can be calculated from    
\be
\rho = -\frac{1}{2 h}\frac{\partial h}{\partial w },
~~|\sigma|^2=\rho^2-det(\frac{\partial \gamma_{AB}}{\partial w})/(4h^2).
\ee
Explicitly we find for the metric (\ref{m22})-(\ref{m33})                       

\be
\rho = -\cot{2w} -\frac{q}{2f\cos{w} }, \label{rhoex}
\ee
\be
|\sigma|^2 = \rho^2 + \frac{k^2 \cos^2{\theta}}{\eta} 
+  \frac{\eta-2q}{ f \sin{w}}. \label{sigma2}
\ee
It is seen that both $\rho$ and $|\sigma|$ diverge at focal points $f=0$. 
The quantity $1/j^2= |\sigma|^2/\rho^2$ measures the anisotropic part 
of distance change to neighbouring null geodesics along a given ray. It can
be written as
\be  \fl           
j^{-2}= 1+ 
\frac{4f(\eta -2q)\sin{w}\cos^2{w}}                    
{(f\cos{2w}+q\sin{w})^2} 
+\frac{4f^2k^2\cos^2{\theta}\sin^2{w}\cos^2{w} }{\eta(f\cos{2w}+q\sin{w})^2}.
\ee
$1/j^2$ evidently goes to 1 for $f \rightarrow 0$, the same limit is reached
 at keel singularities $w=n\pi$.

\subsection{Geometry of caustics}

The affine parameter $w$ gives rise to a foliation of the cone,     
but spacelike surfaces $w=const$ have no 
invariant meaning since $w$ is not uniquely determined.      
There exist however invariantly defined two-surfaces  
on the cone, e.g. the spacelike "zero divergence" surfaces. 
Here $\rho=0$, and from (\ref{rhoex}) one obtains their equation as 

\be
-\frac{\tan{2w}}{2w}=
 \frac{(k^2-1)\sin^2{\theta}}{k^2-1+ (k^2+1)\cos^2{\theta} }.
\ee
Since the {\it rhs} is not negative for the 
metrics considered here, such surfaces can only occur at points where
$\tan{2w}/(2w)$ is negative or null, that is in the range
$(2m-1)\pi/2 \leq 2w \leq m\pi, ~~m=\pm 1,\pm 2, \pm 3 ...$.

Other invariantly defined subsets of the cone are the focal surfaces 
$\cal{F}$ (described as "points of the second kind" in \cite{os62}). 
Their equation $f=0$ can be written as 

\be
-\frac{\tan{w}}{w} 
= \frac{k^2-1}{k^2}\tan^2{\theta}, \label{fock}
\ee
thus focal surfaces occur at points with   
$(2n-1)\pi/2 \leq w \leq n\pi, ~n=\pm 1,\pm 2, \pm 3 ... $,
where $\tan{w}/w$ is negative.
Contrary to the zero-divergence surfaces, focal surfaces are  two-dimensional 
(finite and, as will be argued, non-geodesic) {\it null surfaces}:  
Solving (\ref{fock}) for $\theta$ and introducing this function $\theta_f(w)$ 
in (\ref{teq})-(\ref{zeq}), 
we obtain a parametric representation of $\cal{F}$.
The intrinsic metric of the focal surface                   
follows from (\ref{m22})-(\ref{m33}) as 
\be
ds^2 = f_{ww} dw^2 +2 f_{w\varphi} dw d\varphi +f_{\varphi\varphi} d\varphi^2
\ee 
with 
\ba
f_{ww} &=& \frac{T\sin^2{w}(1+T\cos^2{w})^3}{l^2(k^2-1)^4
\cos^4{w}(1+T)^3(k^2(1+T)-1)},
 \label{f22} \\
f_{w\varphi} &=& 
-\frac{2 T \sin^2{w} (1+T\cos^2{w} )^2} {l^2(k^2-1)^{5/2}(1+T)^{5/2}\cos^2{w}
\sqrt{k^2(1+T)-1)}}, \label{f23} \\
f_{\varphi\varphi} &=&\frac{4T \sin^2{w}(1+T\cos^2{w})}             
 {l^2 (k^2-1)(1+T)^2}                           \label{f33}  
\ea
and $T=-\tan{w}/w$. The range of the coordinates $w,\varphi$ for the $n$th 
focal surface is $0\leq \varphi \leq 2\pi,~(2n-1)\pi/2 \leq w\leq n \pi$.
Since $f_{ww}f_{\varphi\varphi}-f_{w\varphi}^2 = 0$, we have the metric of   
two-dimensional null surfaces, with metric components depending only 
on $w$. 

We consider the first focal surface $n=1$ in more detail.  
$\mathcal{F}_1$ is                              
shown in Fig. 1 as projection into the 3-space 
$t=const$, using $(x=r \cos{\phi},~y=r\sin{\phi},~z)$ as spacetime coordinates. 
\begin{figure}[tbp]
\includegraphics[width=0.3\textwidth]{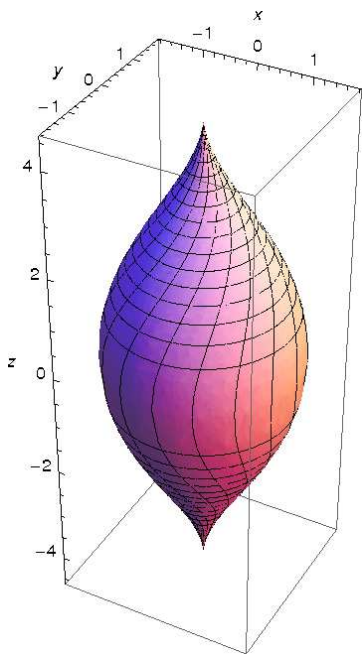}
\caption{ 
The first focal surface  $\mathcal{F}_1$ of the $l=1$, $k^2=2$ (G\"odel) 
light cone 
is shown as projection into the 3-space $t=const$ using 
spacetime coordinates $x=r \cos{\phi},y=r\sin{\phi},z$. The cusps at top and 
bottom lie on the exceptional rays and are the intersection points with the 
first keel. The projected keel is the part of the $z$-axis between top and 
bottom.-  We add a comment of caution. The 3-space $t=const$ with the 
metric tensor $g_{ik}$ is a curved Riemannian $V^3$, while the plot 
must use the Euclidean $R^3$. Thus distances and angles cannot be 
represented correctly. In particular, distortions can occur if the map 
between $V^3$ and $R^3$ becomes singular,                                 
e.g. at points where  
$\det|g_{ik}|= 4(\xi+(1-k^2)\xi^2)/l^2$ 
is zero.  This happens at $\xi = 1/(k^2-1)$, 
corresponding to the "equatorial line" of $\mathcal{F}_1$, the intersection
of $\mathcal{F}_1$ with the plane $z=0$. The plot give the wrong impression
that this line has nonzero length, while the true $V^3$-length is zero, cf. 
(\ref{tl}).}
\hrulefill
\end{figure}
The surface is smooth except at $w=\pi/2$ and $w=\pi$. 
At $w =\pi/2$, $f_{\varphi\varphi}$ tends to zero, $f_{ww}$ 
to infinity, their product $f_{ww}f_{\varphi\varphi}$  (or $f_{w\varphi}^2 $) 
is finite  and equal to the constant $ \pi^4/(4l^4k^2(k^2-1))$. 

We may try to explain this geometrically.  
As noted above, the  sign of the metric component $g_{\phi\phi}$ in 
(\ref{gt0}) decides whether the closed coordinate lines of $\phi$ are 
spacelike or timelike. 
Calculated on the light cone, $g_{\phi\phi}$ becomes 
$\gamma_{33}$, and calculated on a focal surface on the cone, 
$\gamma_{33}$ becomes $f_{\varphi\varphi}$, the $\phi$-coordinate lines
are $\varphi$-coordinate lines on $\mathcal{F}_1$. Contrary to $g_{\phi\phi}$,
$\gamma_{33}$ and $f_{\varphi\varphi}$ cannot become negative, they reach zero
only at isolated points or lines.
Apart from coordinate singularities, 
 the zeros of $\gamma_{33}$  are found at 
keels $w= n\pi$ and as zeros of $q$.                      
$q$ becomes zero only for the $\varphi$-coordinate line at  
$\theta=\pi/2,~w=\pi/2$   (or $w=(2n-1)\pi/2, n=0,
\pm1,2,3...$ at other focal surfaces $\mathcal{F}_n$).  
This  particular line is a closed null line, with zero length also 
from the viewpoint of the cone geometry (it is one of the closed null 
curves on the optical horizon). 
The other $\varphi$-coordinate lines on $\mathcal{F}_1$ (the "parallels" in
Fig. 1) are spacelike for $\pi/2<w<\pi$. Their total length 
\be
L(w)= \int^{2\pi}_0 d\varphi \sqrt{f_{\varphi\varphi}} =
\frac{8\pi\cos{w} \sin^3{w}(\sin{w}\cos{w}-w)}{l^2(k^2-1)(w\cos{w}-\sin{w})^2} \label{tl} 
\ee
increases for $w\geq\pi/2$  from zero to a maximum % at $w=  $ 
and declines to zero for $w\rightarrow \pi$, 
 when the keel is reached. Here $\mathcal{F}_1$ shrinks to the two
cusp points, and all components            
$f_{ww}, f_{w\varphi}, f_{\varphi\varphi} $   vanish. 

Instead of slicing by $\varphi$-lines we can represent $\mathcal{F}_1$
by the lines $ y^i= (w,\theta_f(w), \varphi = const)$, the twisted "meridians"
in Fig.1.
Their tangent vector is  
$\frac{dy^i}{dw}= (1,\frac{d\theta_f}{dw},0)$, with a norm  
given by $(d\theta_f/dw)^2\gamma_{22}=f_{ww} \geq 0$.                  
Thus these lines are spacelike except at the end points 
$w=\pi/2,w=\pi$.  
  
One can construct a focal surface $\cal{F}$ in a still different way. 
A two-dimensional 
null surface always admits a foliation by null lines: 
The equation $f_{AB}f^B=0$ has solutions $f^A$ different from zero since      
$det |f_{AB}| =0$. The tangent lines to these directions can be taken as
$u$-coordinate lines of a new $(u,v)$ coordinate system on $\mathcal F$. 
In $(u,v)$ coordinates the inner metric of a two-dimensional null surface is
represented by the normal form 
\be
ds^2 = F(u,v) dv^2. 
\ee
Explicitly, the transformation from $(w,\varphi)$ to $(u,v)$ is given  for 
$\mathcal F_1$ by
\ba \fl \qquad
  u &=& w, \\ \fl \qquad
  v &=& \varphi - \int^{\pi}_w \frac{dw(w-\sin{w}\cos{w})}
  {\cos{w}\sqrt{\sin{w}-w\cos{w}}\sqrt{k^2(\sin{w}-w\cos{w}) +w\cos{w}}} 
\ea
and the metric function $F(u,v)$ depends only on $u=w$:
\be
F = \frac{4T\sin^2{w}(1+T\cos^2{w})}{l^2(k^2-1)(1+T)^2}.  
\ee
The $u$-coordinate lines on $\mathcal F_1$ have zero lengths  
and can 
therefore be denoted as null lines, but they are different from the null 
geodesic generators of the light cone. From the four-dimensional viewpoint 
they are non-geodesic null curves. It should not be too difficult to develop
a theory of such curves on null hypersurfaces, analogously to Bonnor's
theory  in \cite{bonnor69}.

A way to visualize focal surfaces on the past cone is to locate them on 
the observer sky. We may think of radiation emitted from different parts
of the focal surface.
If we walk down the cone into the past with an  
increasing affine parameter $|w|$, after passing the first zero-divergence 
surface at $|w|=\pi/4$, the first focal surface ($n=1$) 
starts at $|w|=\pi/2$.
Radiation from caustic points at this epoch would appear to the observer 
as luminous ring along the celestial equator $\theta=\pi/2$. For  
larger $|w|$ the focal surface radiation comes in as two luminous parallels, 
moving from the 
equator towards the poles. The pole $(\theta=0)$ and its antipode 
$(\theta=\pi)$ are reached for $|w|=\pi$, marking the two 
singular end points of the focal surface (seen as cusps 
in Fig.1).  
     The cusps are also intersections of the focal surface with 
the two exceptional rays on the                         
past cone (these rays are in the G\"odel case related 
to the rotation direction and
its antipode direction \cite{da06}, and present also for $k^2\neq 2$). 

Keels (denoted as "points of the first kind" in  \cite{os62}) are another 
example of invariantly defined subsets on the cone.
The keels $w=n\pi$, parametrized by $\theta$, are pieces of spacelike
lines connecting the singular end points of the corresponding focal surface. 
 At each keel point labeled with $\theta$ all rays with the same $\theta$ and 
different $\varphi$ intersect. The light cone metric further degenerates 
at keels and becomes a matrix of rank 1, only
$\gamma_{22}= 4f^2/(l^2\eta^3)$ differs in general from zero. At the
two end points,  where 
the keel meets the corresponding focal surface, the matrix rank of 
$\gamma_{ik}$ is
zero, all components $\gamma_{ik}$ vanish here. The keel appears as
second vertex in representations which suppress the $z$-coordinate, 
e.g. in the well-known figure of the G\"odel cone in the Hawking-Ellis 
monograph \cite{hawell73}. But taken all dimensions into account, 
the keel is an extended spacelike 
line, which shrinks to a point only in the $k^2 \rightarrow 1$  limit  of the
G\"odel family.  The invariant length of the  $nth$ keel is given by
\be                     
 \int^\pi_0 d\theta \sqrt{\gamma_{22}}
= \frac{2n\pi \sqrt{k^2-1}}{l}\Bigl(E(\frac{1}{1-k^2}) 
- K(\frac{1}{1-k^2})\Bigr), 
\ee
with $E$ and $K$ as complete elliptic integrals. 

As known from the G\"odel universe or the  lightcone of the
Ozsv\'{a}th-Sch\"ucking anti-Mach metric \cite{os62},
focal surfaces, keels and zero-divergence surfaces  
occur {\it quasi-periodic},
 due to the fact that the
focal function $f(w,\theta)$ is not strictly periodic in $w$, while all other 
functions are {\it circular} functions of the affine parameter.

\subsection{Differential invariants}   

Another quantitative  description of null hypersurfaces                      
is provided by their intrinsic differential invariants.
The quantity $j$ defined as quotient of divergence and shear 
is already an invariant, it is the only invariant depending exclusively 
on the {\it first} derivatives of the cone metric.
A comment is necessary here.
While $|\sigma|$ is always not negative by definition, $\rho$ changes the sign
when the ray passes a caustic (focal surface or keel), and $j$ is $+1$ or $-1$ 
before and behind this point. 
Thus our formal definition produces jumps in $j$ at these points,  
as written down without further explanation in \cite{da06} 
for the G\"odel cone.            
This suggests to redefine the first order invariant 
as   ${\tilde j}=\lambda \rho/|\sigma|,~  \lambda^2 =1$,          
with appropriately chosen $\lambda = f/|f| $  or
 $\lambda = sgn(\rho^2-|\sigma|^2)$, 
to ensure that ${\tilde j}$ is a continuous function through caustics.              
For example, for the Minkowski space null hypersurfaces we have 
${\tilde j}=j_0 +w(1-j_0^2)|\sigma_0|$ as smooth function of $w$,
while $j=\rho/|\sigma|$                                       
shows the unnatural discontinuity at the two focal surfaces.  
Nevertheless we keep $j$ as abbreviation for $\rho/|\sigma|$.            
 
Besides $j$ there exist higher-order invariants   
 \cite{da67}, \cite{da80}, \cite{nur}.    
For their calculation we use a triad formalism \cite{da06}.     
The degenerate inner metric of the cone can be represented by
\be
\gamma_{ik}= t_i\bar{t}_k + \bar{t}_it_k,
\ee
where
$t_i$ is a complex covariant vector intrinsic to the cone. The generator
direction $\epsilon^i$ satisfies $\gamma_{ik}\epsilon^k$=0.
To obtain a complete co- and contravariant triad on the cone we add
further vectors $t^i, \bar{t}^i, \gamma_i$ such that
\be
t_it^i=0,~ t_i\bar{t}^i=1,~ \gamma_it^i=0,~ \gamma_i\epsilon^i=1.
\ee
We use adapted inner cone coordinates
$y^1=w,~y^2=\theta,~y^3=\varphi$ with     $\epsilon^i= \delta^i_1$.
The degenerate metric $\gamma_{ik}$ then reduces to the two-dimensional
metric $\gamma_{AB}$. Comparison with (\ref{m22}-\ref{m33}) gives
(together with $t_1=0$, and up to a rotation)
\ba
t_2 &=& i\frac{f}{l\eta\sqrt{q/2}} 
+\frac{k\cos{\theta}\sin{w}}{l\eta^{3/2}\cos{w} \sqrt{q/2}            }   
(f-q\sin{w}),          \\        
t_3  &=&  \sin{\theta}\sqrt{2q} \sin{w} /(l\eta),   
\ea
The contravariant components $t^i$  are calculated from $t_i=\gamma_{ik}t^k$,
the result is (besides $t^1=0$)  
\ba
t^2 &=& i l\eta\sqrt{2q}/(4f),                        \\
t^3 &=& \frac{l\eta}{4\sin{\theta}\sin{w}\sqrt{q/2}     }  
+i\frac{kl\cos{\theta}\sqrt{2\eta}}{4f\sin{\theta}\cos{w}\sqrt{q}}        
(q\sin{w}-f).                 
\ea
Rotation coefficients related to this triad and of relevance here can
now be obtained from
\ba
\rho + i\nu &=& -t^2\bar{t}_{2,1}- t^3\bar{t}_{3,1}, \label{rco1} \\
\sigma &=&  -\bar{t}^2\bar{t}_{2,1}- \bar{t}^3\bar{t}_{3,1},\label{rco2} \\
\tau &=&  (\bar{t}^2t^3-t^2\bar{t}^3)(\bar{t}_{2,3}- \bar{t}_{3,2}).
\label{rco3}
\ea
One may verify that the expressions for $\rho$ and
$|\sigma|$  obtained from (\ref{rco1},\ref{rco2}) agree with 
(\ref{rhoex}) and (\ref{sigma2}).  
As noted, the components of $t_i, t^i$ are not uniquely determined. This 
affects some rotation coefficients, but not $\rho, |\sigma|$ and also 
not the invariants. The  freedom could (but will not here) be used to 
reach, e.g.,  $\nu = 0$ in (\ref{rco1}). For our choice of the triad   
the real and imaginary part of the complex shear is given by   

\be
 \mathfrak{Re}(\sigma) =
 \frac{2\eta \cos^2{w}- q(1+2\cos^2{w})}{ q\sin{2w}} +\frac{q}{2f\cos{w}},             \label{s1ex}
\ee
\be
 \mathfrak{Im}(\sigma)=  \frac{k(k^2-1)\cos{\theta}\sin^2{\theta}\sin^2{w}}
{q\sqrt{\eta}}. \label{s2ex}                  
\ee

   A null hypersurface has in general four second-order differential 
invariants of the inner geometry, written as complex quantities $I$ and $J$
and conveniently expresssed in terms of rotation coefficients \cite{da67}.  
The quantity $I$ is linear in the second derivatives of the metric,  
with derivatives only along the generators and, like $j$, dimensionless:
\be
I = \frac{i}{|\sigma|}\Bigl(\frac{D\rho}{\rho}
    -\frac{D     \sigma }{     \sigma }\Bigr) 
+ 2 \frac{\nu}{|\sigma|}.    
\ee
Explicitly we find for the real part $I_1$ 
\be
|\sigma|^3I_1 = \frac{2k(k^2-1)\cos{\theta}\sin^2{\theta}  
     (\eta\sin{w}-f)}{f\eta^{3/2} }.
\ee

\noindent The imaginary part $I_2$ has a more complicated 
structure:

\be
|\sigma|^3I_2 = \frac{i_0+i_1f+i_2f^2+i_3f^3}
{2\eta f^2\sin^2{w}\cos{w}(f\cos{2w}+q\sin{w})}.              
\ee
A tedious but straightforward calculation shows that
\be \fl \qquad
i_0 = \eta q^2 \sin^2{w}(2q-\eta),                                    
\ee
\be \fl \qquad
i_1= 2\sin{w}\Bigl( -q^3+\eta q^2(\sin^2{w} -3) + \eta^2 q (5-4\sin^2{w}) 
   -2\eta^3\cos^2{w}\Bigr),
\ee
\be \fl \qquad
i_2 =4 q^2-2q\eta\cos{2w} -\eta^2,
\ee
\be \fl \qquad
i_3 = -2 k^2\cos^2{\theta} \sin{w}.            
\ee
$I_2= Dj/{\rho}$ describes the change of the first-order quantity $j$ 
along the rays.                                                       
$I_1$ is a measure for the rotation of the 
shear directions (i.e. directions 
where the distance change to neighbouring rays is 
a maximum or minimum)
relative to the generator congruence. If $I_1$ is zero (as for 
null hypersurfaces in a Minkowski or conformally related spacetime), 
the shear directions always point to the {\it same} neighbouring null 
rays if one follows a ray.

 The complex invariant $J$ has the dimension $(length)^{-1}$,
 is nonlinear in the second-order derivatives of the inner metric and 
involves additionally transversal derivatives \cite{da67}.
$J$ describes changes 
of the nullsurface geometry in transversal directions,  but is 
considerably more complicated than $I$ and will be discussed 
elsewhere.              

The behaviour of invariants at and in the neighbourhood of focal singularities 
is  of interest. While the rotation coefficients  $\rho$ and $\sigma$ 
show singularities,  the  
invariants tend to have finite values. We have already noted  
$j \rightarrow \pm 1$
 at focal points         
and keels. Expanding $I$ near $f=0$ in powers of $f$ leads to 
 \ba
 I_1 &=&  \frac{16k(k^2-1)\cos{\theta}\sin^2{\theta}\sin{w}\cos^3{w}}
{q^3\sqrt{\eta}}f^2 
     + o(f^3), \\
  I_2 &=& \frac{4\cos^2{w}(\eta-2q)}{q^2\sin{w}}  f + o(f^2).    
 \ea
Remarkably, the second-order differential invariants $I_1,I_2$ 
vanish at focal points.
This also holds at keels $w=n\pi$: 
Writing $w-n\pi=x$, one obtains
for small $x$
\ba
I_1 &=& 16 (-1)^{n+1} k(k^2-1)\eta^{-3/2}\cos{\theta}\sin^2{\theta}~x^3 +o(x^4), \\ 
I_2 &=&   \frac{4(-1)^{n}\eta~x}{n\pi(k^2-1)\sin^2{\theta}}  + o(x^2).   
\ea
For comparison we note that 
null hypersurfaces in a Minkowski spacetime satisfy $I_2= 1/j-j$,        
thus $I_2$ vanishes at caustics, $I_1$ is already zero everywhere.
 
\subsection{Comments on the G\"odel case as treated in \cite{da06} }
For the G\"odel cone ($k^2=2$) some  differential invariants have been 
calculated already in \cite{da06}.  The present paper    
uses different four-dimensional coordinates as well as different 
transversal light cone coordinates $y^A$. 
The latter is motivated by the topological fact that        
no coordinate system can cover the whole sphere without singularity. 
The coordinates $u,v$ in \cite{da06} 
avoid a singularity in the direction of the rotation axes,
they become singular in equator directions instead.   
The polar angles $\theta,\phi$ here avoid the equator singularities 
but show the usual pole singularities. The relation between both systems    
of transversal coordinates  is given by  
$\cos{\phi}=(1-u^2)/(1+u^2),~ \sin{\theta}= \sqrt{2}(v^2-1)/(v^2+1).$
 For $k^2=2$, eqn. (\ref{fock}) thus becomes the focal equation
$ -\tan{w}/w= (v^2-1)^2/(6v^2-1-v^4) $, eqn. (81) in \cite{da06}.-
We note a misprint in eqn. (48) of \cite{da06}: the denominator should
read $f_2+4(1+f_1)$ instead of $f_2$.

\section{The Rebou\c{c}as-Tiomno $G_7$ metric $k^2=1$}
The case $k^2=1$ was excluded so far, we treat it separately.         
 Rebou\c{c}as and Tiomno introduced this special case as "the first 
exact G\"odel-type solution of Einstein's equations describing a completely
causal spacetime-homogeneous rotating universe" \cite{RT1}. The lightcone    
becomes very simple in this model. 
Since the Weyl tensor 
vanishes for $k^2=1$ (see (\ref{weyl})), the spacetime metric is conformal to 
the Minkowski spacetime, thus also the lightcone metric is conformal to the 
Minkowski cone metric. One obtains in the limit $k^2\rightarrow 1,
f \rightarrow \sin{w}\cos^2{\theta}, \eta \rightarrow \cos^2{\theta}, 
q \rightarrow \cos^2{\theta}$  of preceding formulae:
\be
\gamma_{22}= \frac{4\sin^2{w}}{l^2  \cos^2{\theta}},
~\gamma_{23}=0,
~\gamma_{33}= \frac{4\sin^2{w}}{l^2}\tan^2{\theta}.
\ee
The square root $h$ of the determinant $|\gamma_{AB}|$,
\be
h=       4\sin{\theta}\sin^2{w}/(l^2\cos^2{\theta}),  
\ee
vanishes at the points $w = n\pi$ (the only other zeros correspond to the
coordinate singularity).
All light rays from the vertex $w=0$ meet again at the points $w=  n\pi$ 
($n$ integer), which are also vertices. Thus every pair of focal surface 
and keel in the $k^2 > 1 $ family of metrics has collapsed into a 
single vertex in the limit $k^2 \rightarrow 1$.  
The shear of the cone vanishes, only the divergence differs from zero:
\be
\rho = -\cot{w}.
\ee
$\rho$ increases from $-\infty$ at $w=0$ to zero at $w=\pi/2$ and decreases
again until $-\infty$ at the next vertex $w=\pi$.  
The lightcone belongs to a type of
null hypersurfaces characterized by $\rho \neq 0,~|\sigma|=0$ and 
denoted as "type 5" in the classification of \cite{da80}.
There exist no second-order inner differential invariants for this class.

The high symmetry of the Rebou\c{c}as-Tiomno metric is reflected by the
 existence of a 
symmetry group $G_7$ \cite{TRA}, see also \cite{RT2} for further discussions.

\section{Static degeneration $k^2 \rightarrow 0,~ l~finite$ }
The limit $k\rightarrow 0$, keeping $l$ finite, requires $\Omega\rightarrow0$.
It represents the static degeneration of the G\"odel family and has the 
simple line element 
\be
ds^2 = -dt^2+dr^2+\frac{\sinh^2{(lr)}}{l^2}d\theta^2+dz^2. \label{degm}
\ee
Teixera, Rebou\c{c}as  and {\AA}man have shown that this metric admits 
a six-parameter group of motions  $G_6$ \cite{TRA}. The only nonvanishing 
components of the Ricci tensor are $R_{rr}=-l^2$ and $R_{\theta\theta}=
-\sinh^2{(lr)}$ with a constant Ricci scalar $R=-l^2$, and the Riemann tensor
has only one independent nonvanishing component 
$R_{r\theta r\theta}=-\sinh^2{(lr)}$.
Thus the metric cannot easily be interpreted as solution of the field
equations, it is nevertheless interesting geometrically because of its high 
symmetry. 
The null geodesics starting at the origin 
$r=0,~z=0$ are given by (here $w >0$ corresponds to the past cone)
\be t=-w,~ r=w\sin{\theta},~\phi = \phi_0,~ z = w \cos{\theta}.
\ee
The equations show that the cone generators do not re-converge
as generally for the G\"odel family
but extend to null infinity as in the Minkowski spacetime 
(Minkowski is included for $l\rightarrow0$). No focal surface  or keel  
exist for finite $w$. Nevertheless the
cone geometry  and in particular the asymptotic behaviour of the cone  
significantly differ from Minkowski.
The cone metric with the intrinsic 
coordinates $y^1=w,~y^2=\theta, y^3=\varphi$ is given by           

\be
         \gamma_{22}=w^2,~ \gamma_{23} =0,~ \gamma_{33}= 
         \frac{\sinh^2{(l w \sin{\theta})}}{l^2},
\ee
thus the determinant $|\gamma_{AB}|$ is the square of the function 
\be
h = \frac{w}{l}\sinh{(l w \sin{\theta})}.
\ee
Hence, apart from the vertex $w=0$ and the coordinate singularity on the 
symmetry axis, there exist no further singularities on the
cone.  Divergence and shear of the rays follow as
\ba
\rho &=& -\frac{1}{2w}-\frac{l\sin{\theta}}{2}\coth{(l w\sin{\theta})},\\   
\sigma &=& \bar{\sigma}= \frac{1}{2 w} 
-\frac{l\sin{\theta}}{2}\coth{(l w\sin{\theta})}.  
\ea
The divergence increases from $-\infty$ at the vertex $w=0$ to 
$-l\sin{\theta}/2$ for $w\rightarrow\infty$, if one goes down the 
past lightcone, thus it always remains negative. The (real) shear starts with
zero at the vertex, becomes negative for increasing $w$ and reaches the
same negative limit $-l\sin{\theta}/2$ as the divergence for 
$w\rightarrow\infty$. One also has
\be
\rho^2-|\sigma|^2 = \frac{l\sin{\theta}}{w}\coth{(l w\sin{\theta})},
\ee               
which is always positive, thus the lightcone consists exclusively of 
elliptic points.

The second-order invariants $I_1$  and $J$ are zero, but  
$I_2$ is different from zero and  
given by
\be
I_2 = \frac{4l w S\sin{\theta} (CS -l w\sin{\theta})}{(l wC\sin{\theta}+S)
(l w C\sin{\theta}-S)^2} 
\ee
with $S\equiv \sinh{(l w\sin{\theta})},
~C\equiv \cosh{(l w\sin{\theta})}$.   
One verifies that asymptotically, for $ w \rightarrow\infty$, 
$j\rightarrow -1,~ I_2\rightarrow 0$, which are standard limiting values 
at caustics.

\section{Cyclic structure on general null hypersurfaces}

At the end of their pioneering paper \cite{os62}, 
\os ask for the origin of the periodicity structure. 
A certain answer can be given by going back to the
Sachs equations \cite{sachs61}, 
the differential equations governing divergence and shear on a  null 
hypersurface $\cal{N}$ in terms of Ricci and Weyl tensor 
projections into $\cal{N}$:                         
\ba
D\rho &=& \rho^2+\sigma\bar{\sigma} + \omega,  \label{req1}\\
D\sigma &=& 2(\rho - i\nu)\sigma    + \psi, \label{peq}
\ea
with  
\be \fl \qquad
D= p^\mu \partial_\mu= \frac{\partial}{\partial w},~ 
\rho= -p_{\mu;\nu}{\bar t}^\mu t^\nu,~
\nu =  i {\bar t}_{\mu;\nu}{\bar t}^\mu p^\nu,~
\sigma= -p_{\mu;\nu}{\bar t}^\mu {\bar t}^\nu,~
\tau = - {\bar t}_{\mu;\nu}{\bar t}^\mu {\bar t}^\nu,~ \label{rotco}
\ee

\be
 \omega =\frac{1}{2}R_{\mu\nu} p^\mu p^\nu, ~~\psi = 
C_{\mu\nu\rho\sigma}p^\mu{\bar t}^\nu p^\rho{\bar t}^\sigma.\label{op}
\ee
The null vector $p^\mu$ is the direction of the cone generators, 
the complex null vector $t^\mu$ spans spacelike 
directions in $\cal{N}$ orthogonal to $p^\mu$.    
It should be stressed that $\omega$ and $\psi$
- in spite of their origin as projections of four-dimensional 
quantities - depend only on the metric $\gamma_{ik}$, they are 
{\it objects of the inner geometry of $\cal{N}$}.            
The rotation coefficients $\rho,\nu,\sigma,\tau$ defined in (\ref{rotco})
agree with those calculated from (\ref{rco1})-(\ref{rco3}) for
 the G\"odel family. 

The Sachs equations are the first equations to be solved on $\cal{N}$ for a
characteristic initial value problem based on the Newman-Penrose formalism 
and starting from  $\cal{N}$.      
Thus in a sense they can be considered as the Einstein field equations in a 
nutshell, being nonlinear and ruling the influence of matter ($\omega$) 
on the nullsurface geometry. 
The Penrose equations (\ref{mink}) follow as solution  of the Sachs 
equations in the absence of
matter and for a vanishing Weyl tensor 
($\omega=\psi=0$).
For the lightcone of the G\"odel family,   
calculation of $\omega$ and $\psi$ gives
\be
\omega=   2- \frac{k^2\cos^2{\theta}}{(k^2-\sin^2{\theta}}),
\label{o}
\ee
\be
\mathfrak{Re}(\psi) =
 \frac{2(k^2-1)\sin^2{\theta}}{q\eta}(-q+2\eta \cos^2{w}), 
\ee
\be
\mathfrak{Im}(\psi) = 
 \frac{4k(k^2-1)\cos{\theta}\sin^2{\theta}\cos{w}\sin{w}}{q\sqrt{\eta}}.
\ee
The solution  of the Sachs equations with these right-hand-sides     
is given by (\ref{rhoex}) and (\ref{s1ex},\ref{s2ex}). 

From the Sachs equations on can derive a 
differential equation for an area distance $r$ (not to be confused with 
the radial coordinate $r$),  introduced by $Dr = -\rho r$. 
One obtains with $Q=|\sigma|^2 +\omega$ the Jacobi equation     
\be
DDr + Qr  =0. \label{jacobi}
\ee
The caustics are found as zeros of $r$. 
For the G\"odel family from (\ref{o})  $\omega > 0$, hence $Q>0$. 

The existence of cyclic focal features on many null hypersurfaces (not
only cones) may then be considered as property of the Jacobi 
equation, or, more concretely, as property of the function $Q$.
The linear second-order differential equation  (\ref{jacobi})
belongs to the most widely studied equations in applied mathematics. 
Starting with the classical papers by Sturm, Liouville and Kneser in the 
nineteenth century, there exist many    
theorems which indeed prove a cyclic or oscillatory behaviour  
(with arbitrarily large numbers of zeros of $r$)
for certain functions  $Q>0$. For $Q<0$ the solutions are 
non-oscillatory, but this holds also for some $Q>0$,
e.g. for the Minkowski space caustics.  
The precise dependence of the oscillation feature  on properties of $Q$ 
is an open mathematical problem, see \cite{zettl}.

\ack
This work began as part of a collaborative project with M. Abdel-Megied. 
I thank him for the reference to the paper
by Calv\~ao, Soares and Tiomno  \cite{CST} and for many useful discussions. 
I am also grateful to J. {\AA}man and  M.A.H. MacCallum for comments 
on an invariant characterization  of G\"odel-like metrics.

\end{document}